\documentstyle[aps,prl,twocolumn,psfig]{revtex}
\begin{document}
\bibliographystyle{unsrt}

\draft
\title{Breathers on lattices with long range interaction}
\author{S. Flach}
\address{Max-Planck-Institute for the Physics of Complex Systems, 
N\"othnitzer Str. 38,
D-01187 Dresden, Germany 
}
\date{\today}
\maketitle
\begin{abstract}
We analyze the properties of breathers (time periodic spatially
localized solutions) on chains in the presence of algebraically decaying
interactions $1/r^s$. We find that the spatial decay of a breather
shows a crossover from exponential (short distances) to algebraic
(large distances) decay. 
We calculate the crossover
distance as a function of $s$ and the energy of the breather. 
Next we show that the results on energy thresholds obtained
for short range interactions remain valid for $s>3$ and 
that for $s < 3$ (anomalous dispersion at the band edge)
nonzero thresholds occur for cases where
the short range interaction system would yield zero threshold
values. 
\end{abstract}


The understanding of dynamical localization in classical spatially extended
and ordered systems experienced recent considerable progress. Specifically
time-periodic and spatially localized solutions of the classical equations
of motion exist, which are called (discrete) breathers, or intrinsic
localized modes. The attribute discrete stands for the discreteness of
the system, i.e. instead of field equations one typically considers the
dynamics of degrees of freedom ordered on a spatial lattice. 
The lattice
Hamiltonian is invariant under discrete translations in space. The discreteness
of the system produces a cutoff in the wavelength of extended states, and
thus yields a finite upper bound on the spectrum of eigenfrequencies $\Omega_q$
(phonon band) of
small-amplitude plane waves (we assume that usually for small amplitudes 
the Hamiltonian is in leading order a quadratic form of the degrees of 
freedom). If now the equations of motion contain nonlinear terms, the
nonlinearity will in general allow to tune frequencies of periodic orbits
outside of the phonon band, and if all multiples of a given frequency
are outside the phonon band too, there seems to be no further barrier
preventing spatial localization (for reviews see \cite{sa97},\cite{sfcrw98}).

To cope with breathers in lattice dynamics, one has to face the
problem of (i) quantization of breathers, ii) breathers in the presence
of acoustic phonon bands and iii) breathers in the presence of long range
interactions (e.g. in ionic crystals). 
While (i) still lacks a full understanding, the
correspondence between classical breathers and quantum bound states
is believed to be correct 
(\cite{aao70},\cite{oe82},\cite{afko96},\cite{sfvf97},\cite{wbgs98}). 
The case of
acoustic breathers has been studied in one dimension in \cite{lsm97}
and in two dimensions in \cite{fkt97},
where it was shown that the resonance of a zero frequency component
of the breather (static deformation) with the zero of the acoustic
spectrum leads to an algebraically decaying lattice deformation, but
not to a disappearance of the breather. 

As for the case of long range interactions, less 
results are known (but see e.g. \cite{db97},\cite{gmcr97}).
A general proof of existence of breathers in $d$-dimensional 
lattices with algebraically decaying interactions was obtained
in \cite{bm98} with upper bounds for the spatial decay of 
the breather amplitude. Namely for interactions decaying like $1/r^s$
with $r$ being the distance from the breather center and $s$ some
power $s>d$, the breather amplitude is bounded from above by
a power law $a/r^s$ with $a$ being some nonzero constant\footnote{
This is in contrast to results obtained in \cite{gmcr97}
and \cite{sf98}, where for $d=1$ and $s>3$ exponential decay was
obtained. Actually both decay laws hold, see below.
}.
This result leaves us with two questions. First, how can one
obtain contact with the case of short range interaction (basically
$s \rightarrow \infty$) where exponential localization takes place?
And secondly, what is happening to energy thresholds of discrete
breathers in the presence of long range interactions? 
In the case of short range interaction, simple estimates of
the far distance energy of a breather solution yield the correct
predictions for nonzero energy thresholds \cite{fkm97}. Obviously
these estimates would yield zero energy thresholds for all cases
of long range interactions if the far distance energy is calculated
with the help of a spatial decay $a/r^s$ for the breather (the
far distance energy would be simply $\sim a^2$, and would always
tend to zero if $a\rightarrow 0$, see also \cite{sf98}). We will 
resolve these puzzles in the following.

At this stage it is appropriate to fix the class of Hamiltonians
to be considered further:
\begin{equation}
H=\sum_l \left[ \frac{1}{2}P^2_l + 
V(X_l) + \sum_{l'} W_{|l-l'|}(X_l-X_{l'})\right] \;\;. \label{1}
\end{equation}
Here $P_l$ and $X_l$ are canonically conjugated scalar momenta and
displacements of a particle at lattice site $l$. The on site
potential $V(z)=\sum_{\mu=2}^{\infty} \frac{v_{\mu}}{\mu}z^{\mu}$
can be used as a simple way to generate an optical phonon spectrum,
and the interaction $W_{l}(z)=
\sum_{\mu=2}^{\infty} \frac{\phi_{\mu}(l)}{\mu}z^{\mu}$
should incorporate longe range interactions with 
$\phi_2(l)=\frac{C}{2} \frac{1}{l^s}$. For small values of
$P_l$ and $X_l$ the classical Hamiltonian equations of motion
$\dot{X}_l=\frac{\partial H}{\partial P_l}\;,\;
\dot{P}_l=-\frac{\partial H}{\partial X_l}$ can be linearized
in $X_l$. The corresponding eigenvalue problem when solved for
plane waves $X_l(t) \sim {\rm exp}^{{\rm i}(ql-\Omega_q t)}$ is given
by
\begin{equation}
\Omega_q^2 = v_2 + 2C\sum_{m=1}^{\infty}\frac{1}{m^s}(1 - \cos (qm)) \;\;.
\label{2}
\end{equation}
Let us discuss the properties of $E_s(q)=\Omega_q^2 \geq 0$. First $E_s(q)$
is bounded from above for all $s > 1$ and periodic in $q$ with period
$2\pi$. Most important is that $E_s(q)$ is a nonanalytic function in $q$,
i.e. its $\kappa=(s-1)$-st derivative with respect to $q$ is discontinuous
at $q=0$ (when $s$ is noninteger, $(s-1) <\kappa < s$).  
This follows already from the fact that the convergence radius of
(\ref{2}) is zero for nonzero imaginary components in $q$.
Indeed for even integers $s$ one finds \cite{az68} 
$(E_s(q)- v_2) \sim B_s(q/(2\pi))$
for $0 \leq q \leq 2\pi$. Here $B_s(z)$ is the Bernoulli polynomial
of $s$-th order. Consequently at small $q$ the expansion of $E_s(q)$
contains a term $q^{s-1}$ which leads 
together with the periodicity of $E_s(q)$
to the mentioned nonanalyticity. For odd integers $s$ the expansion of $E_s(q)$
contains a term $q^{s-1}\ln (q)$, and for noninteger $s$ a term
$q^{s-1}$ (follows from ${\rm d}^2 E_s(q)/{\rm d}q^2 =
-E_{s-2}(q) + 2C \zeta (s-2)$ with 
$\zeta (z)$ being
the Riemann Zeta function). 
Finally for small q the leading term in the expansion
of $E_s(q)$ is $v_2+C\zeta (s-2) q^2$ for $s > 3$ 
and $v_2+2C a(s) q^{s-1}$ for $1 < s
< 3$ with $a(s)=\int_0^{\infty} (1-\cos x)/x^s {\rm d}x$
\cite{sf98}.
Note that the dispersion at the upper band edge ($q=\pi$) is completely
analytical, and in leading order always proportional to $(q-\pi)^2$. 
Some of these results have been discussed at length in \cite{mmw63}
(see also original references therein).

Now we can turn to the first problem of the spatial decay of
a breather. In order to generate a breather solution we chose
$v_4 \neq 0$ and all other anharmonic terms in $V(z)$ and $W(z)$
being zero. Since we can only simulate finite system sizes $N$, we
use periodic boundary conditions. In that case we have to define
a cutoff length in the interaction which we chose to be $N/2$ (we
will discuss the corresponding corrections to $E_s(q)$ later).
We calculate breather solutions using a Newton algorithm (see
\cite{ma96}, \cite{sfcrw98} for details). The results for
$s=10,20,30$ are shown in Fig.1. 
\begin{figure}[htbp]
\centerline{
\psfig{figure=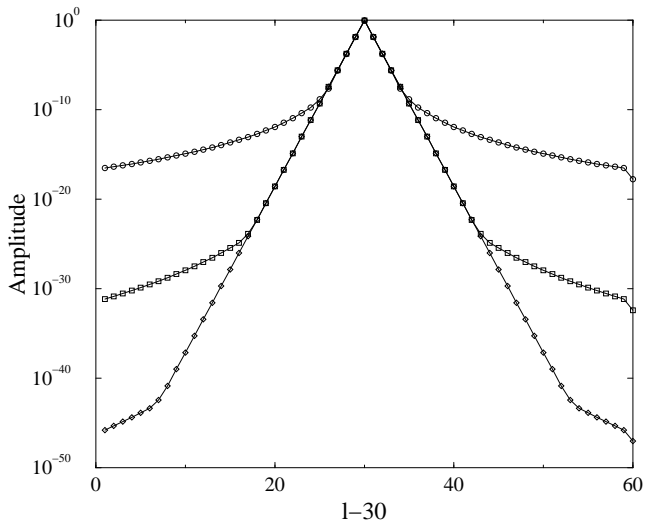,height=200pt,angle=270}
}
\caption[fig1]
{
Breather solution at time $t=0$ with $P_l(t=0)=0$.
The corresponding displacements (amplitudes) $X_l(t=0)$ are plotted
versus lattice site. The nonzero model parameters are
$v_2=v_4=1$, $C=0.01$. The frequency of the solutions is
chosen $T=4.7682$. Circles: $s=10$, squares: $s=20$,
diamonds: $s=30$. Lines are guides to the eye.
}
\label{fig1}
\end{figure}
We observe that
the spatial decay of the breather is {\em exponential} for
small distances from the center, while it becomes {\em algebraic}
(in fact exactly $1/l^s$) after a crossover at some
distance $l_c$ (see Fig.2). 
\begin{figure}[htbp]
\centerline{
\psfig{figure=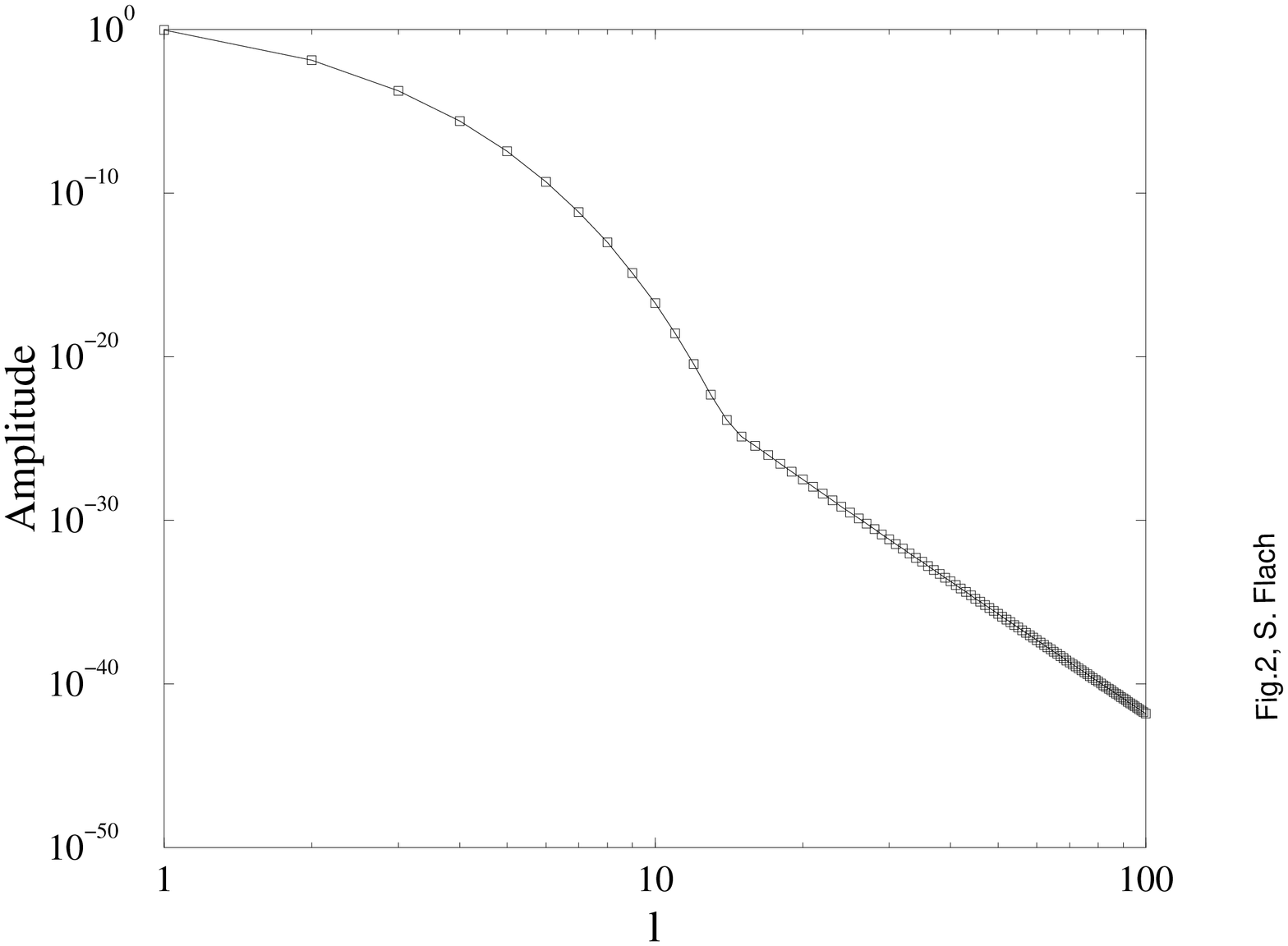,height=200pt,angle=270}
}
\caption[fig2]
{
Same as in Fig.1 but only for $s=30$ in a log-log plot.
}
\label{fig2}
\end{figure}
Note that $l_c$ is $s$-dependent. Moreover, $l_c$
is also depending on the parameter which selects a given breather
solution from its one-parameter family (this parameter could be
the breather frequency, its energy, action or something else). 
In order to understand this result we can proceed along
the following path. Since the breather amplitude decays to
zero with increasing distance from the center, we can linearize
the equations of motion far from the breather center,
keeping the information that we deal with a time-periodic solution
with frequency $\omega_b$ fulfilling the nonresonance condition
$k\omega_b \neq \Omega_q $ (see e.g. \cite{sfcrw98} for details
and also for exceptional 
nonlinear corrections which are however not important at this 
stage). 
Assuming $X_l(t)=\sum_k x_l(k) {\rm e}^{{\rm i}k\omega_b t}$ we 
find linear difference equations for $x_l(k)$ which do not mix
in $k$-space. The spatial decay of the $k$-th amplitude is
then given by the lattice Green's function
\begin{equation}
G_{\lambda}(l) = \int_0^{2\pi} \frac{\cos (ql)}{\Omega_q^2 - \lambda}
{\rm d}q\;\;,\;\;\lambda=k^2\omega_b^2\;\;. \label{3}
\end{equation}
The spatial decay of the breather is thus given by
the convergence properties of the Fourier series, whose coefficients
are given by the rhs in (\ref{3}). As is known, nonanalytic functions
with discontinuities in the $(s-1)$-st derivative (cf. the integrand
on the rhs in (\ref{3})) produce Fourier series which converge algebraically
$1/l^s$ \cite{az68}. From that follows that at large distances
the spatial decay of the breather will be algebraic, which is what
we found in Fig.1.
To obtain the exponential decay at small distances, let us first
slide along the breather family such that the breather frequency
(or one of its multiples) approaches the edge of the phonon band $\Omega_q$.
Then the integrand (\ref{3}) will become very large for wave numbers
close to the band edge which is approached. Applying a stationary
phase approximation to (\ref{3}), i.e. expanding the integrand around
the band edge we obtain
\begin{equation}
G_{\lambda}(l) \sim \int_{-\infty}^{\infty} \frac{\cos (ql)}{
v_2-\lambda + C\zeta (s-2) q^2 } {\rm d}q
\label{4}
\end{equation}
for $s>3$ and
\begin{equation}
G_{\lambda}(l) \sim \int_{0}^{\infty} \frac{\cos (ql)}{
v_2-\lambda + 2Ca(s) q^{s-1}} {\rm d}q
\label{5}
\end{equation}
for $1 < s < 3$. 
Standard evaluation of (\ref{4}) (closing the integration contour
in the complex plane by adding a half circle with infinite radius
and evaluating the residua) yields $G_{\lambda}(l) \sim {\rm e}^{
-\sqrt{v_2-\lambda}l}$ for $s >3$, i.e. exponential decay 
\cite{gmcr97,sf98}! On the
other side, (\ref{5}) yields (closing the integration contour 
in the complex plane by adding a quarter circle and returning
to zero along the positive imaginary axis, and noticing that there are no
poles of the integrand in the enclosed first quadrant including
the imaginary axis) $G_{\lambda}(l) \sim 1/l^s$ for $ 1 < s < 3$
\cite{gmcr97,sf98}.

Now we can explain the observed crossover from exponential to
algebraic decay in Fig.1. Indeed, the stationary phase approximation
for these cases leads to (\ref{4}) in the limit $(v_2-\lambda) \rightarrow
0$. This approximation neglects higher order terms in the expansion
of $E_s(q)$ around $q=0$ which necessarily contain nonanalytic terms.
Consequently (\ref{4}) probes (\ref{3}) for not too large distances
(this is counterintuitive to the assumption that the stationary phase
approximation is correct for large $l$ \cite{gmcr97,sf98}, which it
is not).
Thus we can explain the observed crossover. We can even estimate
the crossover distance $l_c$ using a simple argument. A tagged
site with index $ l < l_c$ and $l > 0$ (the center of the breather 
is located at $l_b=0$) will
experience forces from all other sites with index $l'$ according to (\ref{1}). 
The amplitude of these forces will monotonously decay to zero
for increasing $l'$ with $l' > l$. However the amplitude of 
the forces for decreasing $l'$ will be given by $(l-l')^{-s}{\rm e}^{
\nu ((l-l')}$ for $0 < l' < l$ (here $\nu$ is the given exponent
of the spatial decay for $|l| < l_c$). Since for negative
$l'$ the amplitude of these forces will again monotonously decay to
zero, the worst case is reached when $l'=0$. If this force acting
from the center of the breather on site $l$ is comparable to
the forces acting on $l$ from its nearest neighbours, the exponential
decay will be violated. This condition yields $ l_c^{-s}
{\rm e}^{\nu l_c} = 1$ or
\begin{equation}
\frac{\ln l_c}{l_c} = \frac{\nu}{s} \;\;. \label{6}
\end{equation}
This equation has either two solutions or none. For the larger
(physically relevant) solution we find
$l_c \rightarrow \infty$ if $\nu/s \rightarrow
0$, while the smaller one yields 1 in this limit and
is not of interest. 
Thus for $s>3$ exponential decay is reobtained either for
large $s$ or for breathers with frequencies close to the phonon band
edge. Since we are considering a lattice, the exponential
decay part will disappear if $l_c \approx 1$ or smaller. 
For $s=20$ and $\nu=4.2724$ we obtain $l_c= 11.39$,
and for $s=30$ and the same value of $\nu$ the result is
$l_c=21.56$.
We miss the observed crossovers in Fig.1 by just
two sites. 

For $1 < s < 3$ no exponential decay is observed provided the
breather frequency is located in the gap below the phonon band.
For breather frequencies above the phonon band the dispersion
at the upper band edge yields always quadratic dependence in $q$ 
(see above) and thus there will be always a crossover from
exponential to algebraic decay (provided $l_c > 1$). All these results
were verified by calculating corresponding breather solutions.

To conclude this part we want to stress that a modified
interaction $\phi_2(l) \sim (-1)^l / l^s$ will simply exchange
the notation of upper and lower phonon band edges, and the
case of acoustic interactions is obtained by letting $v_2 \rightarrow 0$.

Let us now turn to the question of energy thresholds for
breathers in the presence of long range interactions. 
There are two lines of argumentation known from the short range
interaction case \cite{fkm97}. The first one estimates the
far distance energy of the breather solution in the limit when
the amplitude of the breather center is small and thus its
frequency is close to a phonon band edge (the only limit where
the breather energy can actually become small). Using exponential
spatial decay the result is that the breather energy tends to
zero only if $v_3 \neq 0$ or/and $v_4 \neq 0$, stays finite
if $v_3=v_4=0$ and $v_5 \neq 0$ or/and $v_6 \neq 0$, and diverges
if $v_3=v_4=v_5=v_6=0$ and $v_{\mu} \neq 0$ for some $\mu \geq 7$
(see \cite{fkm97} for details). In the case of long range interaction
$l_c$ tends to infinity in this limit for $\omega_b > \Omega_q$ or
$\omega_b < \Omega_q$ and $s\geq 3$. Consequently the breather
energy will have the same qualitative behaviour as in the case of short range
interaction (the results are similar to those obtained in 
\cite{fkm97} with the tendency that the height and the
position of the energy minima shift to larger values with
decreasing values of $s$). 

However for $\omega_b < \Omega_q$ and $1 < s < 3$
no exponential decay is observed and the far distance energy
of the breather is given by $ \sim A^2\int \frac{1}{r^{2s}}{\rm d}^dr$
where $A$ is the amplitude of the breather center. This energy
will always vanish in the limit of zero amplitude. However we
are in posess of a second line of argumentation for the behaviour
of the breather energy at small amplitudes. For that we consider
a finite system of $N$ sites. As was shown in \cite{sf96}, 
the band edge plane waves (BEPW) (which can be rigorously defined in the
limit of vanishing amplitudes) undergo tangent bifurcations,
which give birth to discrete breathers. The amplitude $A_c$ of the
BEPW at the bifurcation point (for nonvanishing cubic and/or quartic
terms in the Hamiltonian) was calculated in \cite{sf96}:
\begin{equation}
A_c \sim \sqrt{|\Omega^2_{BEPW} - \Omega^2_{q_1} |} \;\;,
\label{7}
\end{equation}
where $q_1$ denotes the wavevector closest to the band edge wavevector.
Here we consider periodic boundary
conditions and a cutoff in the interaction at one half of the
system size. This cutoff will induce finite size corrections to the
dispersion $\Omega^2_q$ for all $q$ except for the band edge points.
With $q_1=\frac{2\pi}{N}$ this correction amounts to
\begin{equation}
\Delta_{q_1} = \Omega^2_{q_1}(\infty) - \Omega^2_{q_1}(N) =
2C \sum_{m=N/2+1}^{\infty} \frac{1}{m^s} \left( 
1-\cos (\frac{2\pi}{N}m)\right) \;\;.
\label{8}
\end{equation}
Evaluation of (\ref{7}) for $s > 1$ gives
\begin{equation}
\frac{\Delta_{q_1}}{2C} \approx b(s) \left( \frac{2\pi}{N} \right) ^{s-1}
- 2 \left( \frac{2}{N} \right) ^s
\label{9}
\end{equation}
with $b(s) = \int_{\pi}^{\infty}
\frac{1}{x^s}(1-\cos x){\rm d}x$.
Consequently the correct result for (\ref{7}) and $1 < s < 3$ is
$A_c^2 \sim c(s)/ N^{s-1}$ with $c(s) = 
\int_0^{\pi}\frac{1}{x^s}(1-\cos x){\rm d}x$. The total energy 
$E_c \sim N A_c^2$ 
in the bifurcation point  for $1< s < 3$ is finally given by
\begin{equation}
E_c \sim N^{2-s} \;\;.
\label{10}
\end{equation}
This has to be contrasted to the case of short range interactions
in one-dimensional systems which can be obtained from (\ref{10})
by choosing $s=3$ and is $E_c \sim 1/N$. We thus find, that anomalous
dispersion at the band edge $\sim q^{s-1}$ for $1 < s < 3$ even 
further supports the divergence of the breather energy at small
amplitudes, since for cubic and quartic anharmoncities in the
Hamiltonian, for which no divergence in energy is found for 
short range interactions, energy divergence is obtained for
long range interaction with $s < 2$. These results confirm
studies of nonlinear Schr\"odinger chains with long range
interactions, where $s<2$ marks the appearance of two stable
soliton solutions compared to one for $s >2$ \cite{gmcr97}.

Let us discuss the results. First, we numerically confirm that
discrete breathers persist in the case of long range interactions,
even in the case of anomalous dispersions at the band edge. 
Secondly, the spatial decay of breathers is characterized
by a crossover length which separates exponential from 
algebraic decay. 
Third, we show that the existence of energy thresholds for 
breather solutions is supported by long range
interactions, and can take place when short range interactions
(e.g. in one-dimensional systems) are not capable of producing these
thresholds. Thus we can state, that discrete breathers appear
independent of the lattice dimension and survive acoustic and
anomalous dispersions. Discrete breathers have energy thresholds
provided the lattice dimension is large enough (typically $d\geq 2$)
or the interaction is long ranged (for $d=1$ $s < 2$).

Let us speculate on the value of these energy thresholds for
the lattice dynamics of crystals. As there
is no small parameter in the system, these threshold energies will
be comparable to the energy of a vacancy. Consequently discrete
breathers in two- or three-dimensional crystal lattices will
be high energy excitations, which could play a role close 
to the melting transition. In contrast for one-dimensional 
systems there are no energy thresholds (except for $s < 2$) and
breathers can play a role in the dynamics of molecules and similar
objects also at low temperatures. 
\\
\\
Acknowledgements. \\
I wish to thank D. Bonart and J. B. Page for many discussions which initiated
these studies, C. Baesens and R. S. MacKay 
for sending me their work prior publication (and drawing my attention
to the helpful reference \cite{az68}) 
and Yu. Gaididei, M. Katsnelson, Yu. Kosevich
and O. Yevtushenko
for helpful discussions.



\begin{thebibliography}{10}

\bibitem{sa97}
S.~Aubry.
\newblock {\em Physica D}, 103:201, 1997.

\bibitem{sfcrw98}
S.~Flach and C.~R. Willis.
\newblock {\em Phys. Rep.}, 295:181, 1998.

\bibitem{aao70}
A.~A. Ovchinnikov.
\newblock {\em Zh. Eksp. Teor. Fiz. / Soviet Physics JETP}, 57 / 30:263 / 147,
  1969 / 1970.

\bibitem{oe82}
A.~A. Ovchinnikov and H.~S. Erikhman.
\newblock {\em Uspekhi Fiz. Nauk (russian)}, 138:289, 1982.

\bibitem{afko96}
S.~Aubry, S.~Flach, K.~Kladko, and E.~Olbrich.
\newblock {\em Phys. Rev. Lett.}, 76:1607, 1996.

\bibitem{sfvf97}
S.~Flach and V.~Fleurov.
\newblock {\em J. Phys.: Cond. Mat.}, 9:7039, 1997.

\bibitem{wbgs98}
W.~Z. Wang, A.~R. Bishop, J.~T. Gammel, and R.~N. Silver.
\newblock {\em Phys. Rev. Lett.}, 80:3284, 1998.

\bibitem{lsm97}
R.~Livi, M.~Spicci, and R.~S. MacKay.
\newblock {\em preprint}, 1997.

\bibitem{fkt97}
S.~Flach, K.~Kladko, and S.~Takeno.
\newblock {\em Phys. Rev. Lett.}, 79:4838, 1997.

\bibitem{db97}
D.~Bonart.
\newblock {\em Physics Letters}, A231:201, 1997.

\bibitem{bm98}
C.~Baesens and R.~S. MacKay.
\newblock {\em Preprint}, 1998.

\bibitem{gmcr97}
Yu.~B. Gaididei, S.~F. Mingaleev, P.~L. Christiansen, and K.~O. Rasmussen.
\newblock {\em Phys. Rev.}, E55:6141, 1997.

\bibitem{sf98}
S.~Flach.
\newblock {\em Physica D}, 113:184, 1998.

\bibitem{fkm97}
S.~Flach, K.~Kladko, and R.~S. MacKay.
\newblock {\em Phys. Rev. Lett.}, 78:1207, 1997.

\bibitem{az68}
A.~Zygmund.
\newblock {\em Trigonometric Series}.
\newblock Cambridge University Press, 1968.

\bibitem{mmw63}
A.~A. Maradudin, E.~W. Montroll, and G.~H. Weiss.
\newblock {\em Theory of Lattice Dynamics in the Harmonic Approximation}.
\newblock Academic Press, New York, 1963.

\bibitem{ma96}
J.~L. Marin and S.~Aubry.
\newblock {\em Nonlinearity}, 9:1501, 1996.

\bibitem{sf96}
S.~Flach.
\newblock {\em Physica}, D91:223, 1996.

\end{thebibliography}
\end{document}